\documentclass[a4paper,12pt]{article}

\usepackage{amsmath,amssymb,amsfonts,amsxtra,makeidx,graphics,graphicx,youngtab,amsthm,epstopdf,epsfig}
\usepackage[pdftex]{hyperref}
\usepackage[all]{xy}

\newsavebox{\ns}
\newsavebox{\dbrane}
\newsavebox{\dbshort}

\newcommand{\be}{\begin{equation}}
\newcommand{\ee}{\end{equation}}
\newcommand{\beq}{\begin{equation}}
\newcommand{\eeq}{\end{equation}}
\newcommand{\ba}{\begin{array}}
\newcommand{\ea}{\end{array}}
\newcommand{\bea}{\begin{eqnarray}}
\newcommand{\eea}{\end{eqnarray}}
\newcommand{\ben}{\begin{enumerate}}
\newcommand{\een}{\end{enumerate}}
\newcommand{\bean}{\begin{eqnarray*}}
\newcommand{\eean}{\end{eqnarray*}}
\newcommand{\eref}[1]{(\ref{#1})}

\newcommand{\nn}{\nonumber}

\newcommand{\BC}{\mathbb{C}}
\newcommand{\comment}[1]{}

\newcommand{\CM}{{\cal M}}

\newcommand{\CB}{{\cal B}}

\newcommand{\IC}{\mathbb{C}}

\newcommand{\ud}{\mathrm{d}}

\newcommand{\PE}{\mathrm{PE}}
\newcommand{\PL}{\mathrm{PL}}

\newtheorem{theorem}{\bf Theorem}

\newtheorem{observation}[theorem]{\bf Observation}

\newcommand{\setall}{\setcounter{equation}{0}
        \setcounter{theorem}{0}}

\begin{document}
\pagestyle{plain}
\setcounter{page}{1}
\newcounter{bean}
\baselineskip16pt

\begin{titlepage}

\begin{center}
\today
\hfill Imperial/TP/08/AH/04 \\

\vskip 1.5 cm
{\Large \bf Counting Gauge Invariant Operators in SQCD with Classical Gauge Groups}
\vskip 1 cm 
{\large Amihay Hanany and Noppadol Mekareeya}\\
\vskip 1cm
{\sl Theoretical Physics Group, 
The Blackett Laboratory, \\ Imperial College London, Prince Consort Road,\\ London,  SW7 2AZ,  U.K.\\ ${}$ \\} 
{\tt a.hanany, n.mekareeya07@imperial.ac.uk}

\end{center}

\vskip 0.5 cm
\begin{abstract}
\noindent We use the plethystic programme and the Molien--Weyl fomula to compute generating functions, or Hilbert Series, which count gauge invariant operators in SQCD with the $SO$ and $Sp$ gauge groups.  The character expansion technique indicates how the global symmetries are encoded in the generating functions.  We obtain the full character expansion for each theory with arbitrary numbers of colours and flavours.  We study the orientifold action on SQCD with the $SU$ gauge group and examine how it gives rise to SQCD with the $SO$ and $Sp$ gauge groups.  We establish that the classical moduli space of SQCD is not only irreducible, but is also an affine Calabi--Yau cone over a weighted projective variety.
\end{abstract}
\end{titlepage}

\setcounter{footnote}{0}
\renewcommand{\thefootnote}{\arabic{footnote}}

\newpage
\tableofcontents

\newpage

\section{Introduction and Summary} 
As pointed out in \cite{Gray}, the vacuum moduli space of Supersymmetric Quantum Chromodynamics (SQCD) has a very rich structure from which we can employ various algebraic and geometrical techniques \comment{\cite{Gray:2005sr, Gray:2006jb, Gray:2008zs, fluxcomp}}to gain physical insights.  The plethystic programme, Molien--Weyl formula and character expansion techniques provide a very satisfactory way in constructing generating functions (Hilbert Series) which solve the complicated problem on counting gauge invariant operators.  Having studied the $SU(N_c)$ SQCD in \cite{Gray}, we extend our work to the other classical gauge groups, namely $SO(N_c)$ and $Sp(N_c)$.  Several aspects of the $SO$ and $Sp$ gauge theories, \emph{e.g.} dualities, deconfinement, s-confinement, have been extensively studied in a series of works \cite{Intriligator:1995id, Terning:2003th, Terning:2006bq, Dotti:1998rv, Intriligator:1995ne, Seiberg:1994pq, Intriligator:1995ff, Csaki:1996eu, Luty:1996cg, Leigh:1995qp, Intriligator:1995ax, Csaki:1996sm, Csaki:1996zb}.  

In this paper, we shall focus on $\mathcal{N} =1$ SQCD with $SO(N_c)$ and $Sp(N_c)$ gauge groups with $N_f$ flavours of quarks transforming, respectively, in the vector and fundamental representations of the gauge group.  The global symmetries of the theory are respectively $SU(N_f) \times U(1)_R$ and $SU(2N_f) \times U(1)_R$.   We shall concentrate our attention on the case with a vanishing superpotential.  The vacuum space is conveniently described by
polynomial equations written in terms of variables which are the holomorphic gauge invariant operators (GIOs) of the theory, namely the mesons and baryons for the $SO$ theories, and the mesons for the $Sp$ theories.

To facilitate the reading of this paper, we have highlighted the key
points in bold font as {\bf Observations}.  Below, we collect the main
results of our work.

\paragraph{Outline and Key Points:}
\begin{itemize}
\item In Section \ref{so}, we examine the classical moduli space of $SO(N_c)$ SQCD with $N_f$ flavours.  For $N_f < N_c$,     
the moduli space is $\mathbb{C}^{\frac{1}{2}N_f(N_f+1)}$ (Observation \ref{modSOnf<nc}) with the Hilbert Series given by \eref{gNf<NcSO}.  For $N_f = N_c$, the moduli space is a complete intersection and is, in fact, a single hypersurface in $\mathbb{C}^{\frac{1}{2}N_f(N_f+1)+1}$ (Observation \ref{cinfnc}) with the Hilbert Series given by \eref{HSnfnc}.
For $N_f > N_c$, the moduli space is a non-complete intersection of polynomial relations (syzygies) amongst the GIOs.  We use the plethystic exponential and Molien--Weyl formula to derive generating functions for various $N_f$ and $N_c$.  We also use the plethystic logarithm to count basic generators of GIOs and basic constraints between them.

\item In Section \ref{charexpso}, we synthesise our prior results using representation theory and the character expansion.  It proves useful to write the Hilbert series in terms of characters. This permits the generalisation of our results to an \emph{arbitrary number of colours and flavours}. Subsequently, we obtain an important result, namely the full character expansion of the generating function for \emph{any} $N_f$ in an \emph{arbitrary} $SO(N_c)$ theory (Equations  \eref{charexpNf<Nc}, \eref{gencharexp}, \eref{charexpNf=Nc}).  

\item In Section \ref{sp}, we investigate the classical moduli space of $Sp(N_c)$ SQCD with $N_f$ flavours.  For $N_f \leq N_c$,     
the moduli space is $\mathbb{C}^{\frac{1}{2}N_f(N_f+1)}$ (Observation \ref{modSpnf<nc}) with the Hilbert Series given by \eref{gNf<NcSp}.  For $N_f = N_c + 1$, the moduli space is a complete intersection and is, in fact, a single hypersurface in $\mathbb{C}^{(2N_c +1)(N_c+1)}$ (Observation \ref{ciSp}) with the Hilbert Series given by \eref{nf=nc+1}.
For $N_f > N_c +1 $, the moduli space is a non-complete intersection of syzygies amongst the GIOs.  The plethystic exponential and Molien--Weyl formula are used to derive generating functions for various $N_f$ and $N_c$.  We also count basic generators of GIOs and basic constraints using the plethystic logarithm.

\item The full character expansion of the generating function for \emph{any} $N_f$ in an \emph{arbitrary} $Sp(N_c)$ theory is given in \eref{genSp}.

\item In Section \ref{orientifold}, we study how the $SO$ and $Sp$ gauge theories arise from the $SU$ gauge theory due to an orientifold $\mathbb{Z}_2$ action.  Without specifying the explicit brane construction, we consider  an orientifold projection on the global symmetry, the basic generators, and the basic constraints in the $SU$ theory.  We find that the projection occurs in two steps:  The antifundamental index is first turned into a fundamental index, and the resulting symmetry then gets respectively symmetrised and antisymmetrised in the $SO$ and $Sp$ theories.

\item  In Section \ref{apercu}, we take a geometric aper\c{c}u of the moduli space of SQCD.  We establish that the classical moduli space is an irreducible affine Calabi--Yau cone.

\end{itemize}

\noindent \paragraph{Notation for irreducible representations.} We may represent an irreducible representation of a classical group $G$ by a Young diagram. Let $\lambda_i$ be the length of the $i$-th row ($1 \leq i \leq r \equiv\mathrm{rank}~G)$ and let $a_i = \lambda_i - \lambda_{i+1}$ be the differences of lengths of rows.  Henceforth, we denote such a representation by the notation $[a_1, a_2, \ldots, a_{r}]$.  We denote by $[1,0,\ldots,0]$ the fundamental representations of $SU$ and $Sp$ and the vector representation of $SO$.  We also use the subscripts ${k;L}$ and $k;R$ to indicate respectively the $k$th-postitions from the left and the right, \emph{e.g.} $1_{k;L}$ in $[0, ..., 0, 1_{k ;L}, 0, ..., 0]$ denotes the 1 in the $k$-th position from the left.  For representations of the product group $G_1 \times G_2$, we use the notation $[\ldots;\ldots]$ where the tuple to the left of the `;' is the representation of $G_1$, and the tuple to the right of the `;' is the representation of $G_2$.

\section{$SO(N_c)$ SQCD with $N_f$ flavours} \label{so}  \setall
We specify SQCD with gauge group $SO(N_c)$ and $N_f$ flavours by the
ordered pair $(N_f, SO(N_c))$.  This theory has quarks $Q^i_a$, with flavour indices $i = 1, \ldots, N_f$ and
colour indices $a = 1, \ldots, N_c$.  Thus, there is a total of
$N_cN_f$ chiral degrees of freedom from the quarks. Their quantum numbers are summarised in Table \ref{table:so}.  The excellent reviews collecting this work are \cite{Intriligator:1995id, Terning:2003th, Terning:2006bq}.  For $N_f < N_c$,     
the moduli space is $\mathbb{C}^{\frac{1}{2}N_f(N_f+1)}$ (Observation \ref{modSOnf<nc}) with the Hilbert Series given by \eref{gNf<NcSO}.

\begin{table}[htdp]
\begin{center}
\begin{tabular}{|c||c|ccc|}
\hline
 &Gauge symmetry&{}&Global symmetry&  \\
& $SO(N_c)$ & $SU(N_f)$ &$U(1)_B$ &$U(1)_R$ \\
\hline \hline
$Q^i_a$ & $\tiny\yng(1)$ & $\tiny\yng(1)$ & 1& $\frac{N_f+2-N_c}{N_f}$ \\
\hline
\end{tabular}
\end{center}
\caption{{\sf The gauge and global symmetries of $SO(N_c)$ SQCD with $N_f$ flavours and the quantum numbers of the chiral supermultiplets.}}
\label{table:so}
\end{table}

For $N_f \leq N_c-2$, at a generic point in the classical moduli space, the $SO(N_c)$ gauge symmetry is broken to $SO(N_c - N_f)$.  Since the dimension of $SO(N)$ is $\frac{1}{2}N (N-1)$, there are
\beq
\frac{1}{2}N_c (N_c-1) - \frac{1}{2}(N_c - N_f) (N_c - N_f -1) = N_cN_f - \frac{1}{2}N_f(N_f+1) \nn
\eeq 
broken generators.   Therefore, of the original $N_cN_f$ chiral supermultiplets, only
\beq
N_cN_f - \left[ N_cN_f - \frac{1}{2}N_f(N_f+1) \right] = \frac{1}{2}N_f(N_f+1) \nn
\eeq
singlets are left massless.  Hence, the dimension of the moduli space of vacua is
\beq \label{dimnflnc}
\dim\left( \CM_{N_f \leq N_c-2} \right) =  \frac{1}{2}N_f(N_f+1)~.
\eeq

For $N_f \geq N_c-1$, at a generic point in the moduli space,  the $SO(N_c)$ gauge symmetry is broken completely and hence the number of remaining massless chiral supermultiplets ({\em i.e.}\ the dimension of the moduli space) is given by
\begin{equation} \label{vevSO}
\dim \left( \mathcal{M}_{N_f \geq N_c-1} \right) = N_f N_c - \frac{1}{2}N_c (N_c-1) ~.
\end{equation}

According to \cite{Intriligator:1995id}, we see that the `D-flatness' constraints force us to consider matter field solutions in two cases, namely $N_f < N_c$ and $N_f \geq N_c$.  We shall focus on GIOs in each of these cases below. 

\subsection{The Case of $N_f < N_c$}
We can describe the $\frac{1}{2}N_f(N_f+1)$ light degrees of freedom in a gauge invariant way by the mesons:
\beq
\ba{ll}
M^{i j} = Q^i_a Q^j_b \delta^{ab} & \qquad \mbox{(meson)} ~.
\ea
\eeq
We emphasise that the indices $i$ and $j$ are symmetric.   Therefore, the meson transforms in the global $SU(N_f)$ representation $\mathrm{Sym}^2 [1, 0, \ldots, 0] = [2, 0, \ldots, 0]$.  We note that for the $N_f < N_c$ theory, there are no relations (constraints) between mesons.  Phrasing this geometrically, and noting the dimension from \eref{dimnflnc}, we have that 
\begin{observation} \label{modSOnf<nc}
The moduli space  $\CM_{N_f < N_c}$ is freely generated: there are no relations among the generators.
The space $\mathcal{M}_{N_f < N_c}$ is, in fact, nothing but $\IC^{\frac{1}{2}N_f(N_f+1)}$.
\end{observation}
\noindent Using the plethystic programme, we can immediately write down the generating function of GIOs for $N_f < N_c$ as
\beq \label{gNf<NcSO}
g^{N_f < N_c} (t) = \PE \left[ t^2 \dim~[2, 0, \ldots, 0]  \right] = \PE \left[ \frac{1}{2}N_f(N_f+1) t^2 \right] = \frac{1}{(1-t^2)^{\frac{1}{2}N_f(N_f+1)}}~,
\eeq
where $t$ is a chemical potential which can be taken to be conjugate to the $R$ charge.
We emphasise that this formula does not depend on the number of colours $N_c$. This expression is simply the Hilbert series for $\IC^{\frac{1}{2}N_f(N_f+1)}$, with weight 2 for each meson.

\subsection{The Case of $N_f \geq N_c $}
We can describe the light degrees of freedom in a gauge invariant way by the following basic generators:
\beq\ba{ll}
M^{i j} = Q^i_a Q^j_b \delta^{ab} & \qquad \mbox{(mesons)} ~; \\
B^{i_1 \ldots i_{N_c}} = Q^{i_1}_{a_1} \ldots Q^{i_{N_c}}_{a_{N_c}} \epsilon^{a_1\ldots a_{N_c}} & \qquad \mbox{(baryons)} ~. \\
\ea
\label{gioSQCD}\eeq
For $N_f \ge N_c$, under the global $SU(N_f)$ symmetry,
the mesons $M$ transform in the $[2,0,\ldots,0]$ representation and the baryons $B$ transform respectively in $[0,0,\ldots, 1_{N_c;L},0, \ldots, 0]$, where $1_{j;L}$ denotes a $1$ in the $j$-th position from the left.  The dimensions of these representations are respectively $\frac{1}{2}N_f(N_f +1)$ and $N_f \choose N_c$.

We emphasise that the basic generators in \eref{gioSQCD} are not
independent, but they are subject to the following constraints. 
Any product of $M$'s and $B$'s antisymmetrised on $N_c+1$ (or more) 
upper or lower flavour indices must vanish:
\begin{equation}\label{cons1}
M \cdot *B = 0~,
\end{equation}
where $(*B)_{i_{N_c+1} \ldots i_{N_f}} = \frac{1} {N_c !} \epsilon_{i_1 \ldots i_{N_f}} B^{i_1 \ldots i_{N_c}}$ and a `$\cdot$' denotes a contraction of an upper with a lower flavour index.  We note that this constraint transform in the global $SU(N_f)$ representation $[1, 0, ..., 0, 1_{N_c+1;L}, 0, ..., 0]$. Another 
constraint follows from the facts that the rank of the meson $M$ is $N_c$ and that the product of two epsilon tensors can be written as the antisymmetrised sum of Kronecker deltas:
\begin{equation}\label{cons2}
 \CB^2 = \widetilde{\det} ~M~,
\eeq
where $\CB \equiv B^{1 \ldots N_c} = \frac{1} {N_c !} \epsilon_{i_1 \ldots i_{N_c}} B^{i_1 \ldots i_{N_c}}$ and $\widetilde{\det}$ denotes the product of the non-zero eigenvalues.  We note that this constraint transforms in the global $SU(N_f)$ representation $\mathrm{Sym}^2 [0, \ldots, 0, 1_{N_c;L}, 0, \ldots, 0]$. 

Because of these constraints, the spaces $\CM_{N_f \ge N_c}$ are not
freely generated. Moreover, they also prevent us from writing a generating
function as directly as in \eref{gNf<NcSO}. Nevertheless, we
will see that the Molien--Weyl formula gives us the right answer.

\subsubsection{The Case of $N_f=N_c$}
The special case of $N_f=N_c$ deserves some special attention. 
The total number of basic generators for the GIOs,
coming from the two contributions in \eref{gioSQCD}, is $\frac{1}{2}N_f(N_f+1) +1$.
From (\ref{vevSO}), the dimension of the moduli space is 
\beq \dim
\left( \mathcal{M}_{N_f = N_c} \right) = \frac{1}{2}N_f(N_f+1) ~. \label{consnfnc}
\eeq 
There is one constraint (\ref{cons2}), which in this case can be
reduced to a single hypersurface:
\beq\label{nf=nc}
\CB^2 = \det(M) \ .
\eeq
This constraint transforms in the trivial representation $[0, \ldots, 0]$ of the global symmetry $SU(N_f)$ (as the length of the weight is the rank of $SU(N_f)$ or $N_f-1$, there are no 1's).   Note that the relation \eref{cons1} does not provide any additional information and (\ref{cons2}) constitutes the only constraint.
Since, in this case, the dimension of the moduli
space equals the number of the basic generators minus the number of
constraints, we arrive at another important conclusion:
\begin{observation}\label{cinfnc}
The moduli space $\mathcal{M}_{N_f=N_c}$ is a complete intersection.
It is in fact a single hypersurface in $\IC^{ \frac{1}{2}N_f(N_f+1)+1}$.
\end{observation}

An interesting question to consider is to determine the number of
independent GIOs that can be constructed from the basic generators
(\ref{gioSQCD}) subject to the constraints \eref{cons1} and
\eref{cons2}.  In the case $N_f = N_c$, where the only constraint is
\eref{nf=nc}, the generating function can be easily computed from the knowledge that the modul space is a complete intersection (See \cite{BFHH} for a detailed discussion on this). There are $\frac{1}{2}N_f(N_f +1)=\frac{1}{2}N_c(N_c +1)$ mesonic generators of weight $t^2$ and one baryonic generator of weight $t^{N_c}$, subject to a relation of weight $t^{2N_c}$.  As a result, the generating function takes the form
\bea\label{HSnfnc}
g^{N_f=N_c}(t) &=& \left(1-t^{2N_c}\right) \PE \left[ t^2\: \dim\: [2,0, \ldots, 0] +  t^{N_c} \: \dim \: [0, \ldots,0] \right] \nn\\
&=& \left(1-t^{2N_c}\right) \PE \left[ \frac{1}{2}N_c(N_c +1) t^2 +  t^{N_c} \right] \nn \\
&=& \frac{1-t^{2N_c}}{(1-t^2)^{\frac{1}{2}N_c(N_c +1)}(1-t^{N_c})} = \frac{1+t^{N_c}}{(1-t^2)^{\frac{1}{2}N_c(N_c +1)}}~.
\eea
\noindent This is indeed the Hilbert series of the hypersurface \eref{nf=nc}.

\subsection{Counting Gauge Invariants: the Plethystic Exponential and Molien--Weyl formula}
We have seen that the chiral GIOs are symmetric functions of quarks, which in turn transform in the bifundamental $[1,0, \ldots, 0;1,0, \ldots, 0]$ of $SU(N_f) \times SO(N_c)$.    Since a special orthogonal group falls into one of the two categories of the classical groups, namely $B_{n} = SO(2n+1)$ and $D_{n}= SO(2n)$, we use this notation throughout the section unless indicated otherwise.  We note that the Lie algebras of $B_{n}$ and $D_{n}$ both have the same rank $n$.

To write down explicit formulae and for performing computations we need to introduce weights for the different elements in the maximal torus of the different groups. We use $z_a$ (where $a$ runs over $1,\ldots, n$) for colour weights and $t_i$ (where $i=1,\ldots,N_f$) for flavour weights. These weights have the interpretation of chemical potentials\footnote{
Strictly speaking, they are \emph{not} true chemical potentials conjugate to the number of charges.
They are in fact \emph{fugacities}.
We shall however slightly abuse the terminology by calling them chemical potentials.} for the charges they count and the characters of the representations are functions of these variables.
Correspondingly, the character for a quark is $\chi^{SU(N_f)\times B_{n}}_{[1,0, \ldots, 0; 1,0,\ldots, 0]} (t_i, z_a)$ or $\chi^{SU(N_f)\times D_{n}}_{[1,0, \ldots, 0; 1,0,\ldots, 0]} (t_i, z_a)$ depending on which gauge group we are dealing with.  We further introduce a chemical potential which counts the number of quarks, $t$.
Recall from  \cite{Gray,BFHH,feng,forcella,Butti:2007jv,hanany, Noma:2006pe, Balasubramanian:2007hu, Dolan:2007rq} that a convenient combinatorial tool which constructs symmetric products of representations is the {\bf plethystic exponential}, which is a generator for symmetrisation. To briefly remind the reader, the plethystic exponential, $\PE$,  of a function $g (t_1, \ldots, t_n)$ is defined to be $\exp\left( \sum\limits_{k=1}^\infty\frac{g (t_1^k, \ldots, t_n^k)}{k}\right)$.
Whence, we have that
\beq\label{PEge}\nonumber
\mathrm{PE}\: \left[ t \chi^{SU(N_f)\times B_{n}, D_{n}}_{[1,0, \ldots, 0; 1,0, \ldots, 0]} (t_i, z_a) \right] 
\equiv \exp \left[ \sum\limits_{k=0}^\infty \frac{1}{k}\left( t^k \chi^{SU(N_f)\times B_{n}, D_{n}}_{[1,0, \ldots, 0; 1,0,\ldots, 0]} (t_i^k, z_a^k) \right) \right] ~.
\eeq
A somewhat more explicit form for the character can be
\beq \label{refine}
t \chi^{SU(N_f) \times B_{n}, D_{n}}_{[1,0, \ldots, 0; 1,0,\ldots, 0]} (t_i, z_a) = \chi^{B_{n}, D_{n}}_{[1,0, \ldots,0]}(z_a) \sum_{i=1}^{N_f} t_i~ ,
\eeq
which then gives
\beq\label{PEgen}
\mathrm{PE}\: \left[ \chi^{B_{n}, D_{n}}_{[1,0, \ldots,0]}(z_a) \sum_{i=1}^{N_f} t_i  \right] 
= \exp \left[ \sum\limits_{k=0}^\infty \frac{1}{k} \left(  \chi^{B_{n}, D_{n}}_{[1,0, \ldots,0]}(z_a^k) \sum\limits_{i=1}^{N_f} t_i^k \right) \right] ~.
\eeq
Here, the dummy variables $t_i$ are the chemical potentials
associated to quarks counting the $U(1)$-charges in the maximal torus of the global symmetry.
Henceforth, we shall take their values to be such that $|t_i| <1$ for all $i$.

We emphasize that in order to obtain the generating function that counts \emph{gauge invariant} quantities, we need to project the representations of the gauge group generated by the plethystic exponential onto the trivial subrepresentation, which consists of the quantities \emph{invariant} under the action of the gauge group.
Using knowledge from representation theory, this can be done by integrating over the whole group (see, \emph{e.g.}, \cite{pouliot, romelsberger, Aharony:2003sx, Heckenberger:2007ry}).
Hence, the generating functions for the $(N_f, B_{n})$ and $(N_f, D_{n})$ theories are given by
\beq \label{genfn}
g^{(N_f, B_{n})}, g^{(N_f, D_{n})} = \int_{B_{n}, D_{n}} \ud \mu_{B_{n}, D_{n}}\: \mathrm{PE}\: \left[ \chi^{B_{n}, D_{n}}_{[1,0, \ldots,0]}(z_a) \sum_{i=1}^{N_f} t_i  \right] ~.
\eeq
These formulae are the {\bf Molien--Weyl formulae}.

Let us write the above generating function in a ready-to-calculate form.  For each category, we take a basis for the dual space of the Cartan subalgebra to be $\{ L_a \}_{a=1}^{n}$. 
A convenient choice which we shall adopt is  $L_a = (0, \ldots,0, 1_{a;L},0, \ldots, 0)$, where the length of the tuple is $n$.  The weights of the fundamental representations of $B_{n}$ and $D_{n}$ are respectively $\{0, \pm L_a \}$ and $\{\pm L_a\}$.  With this choice, we can write down the characters of the fundamental representations of $B_{n}$ and $D_{n}$ respectively as
\bea
 \chi^{B_{n}}_{[1,0, \ldots,0]} (z_a) &=& 1+ \sum_{a=1}^{n} \left( z_a + \frac{1}{z_a} \right) ~, \nn \\
 \chi^{D_{n}}_{[1,0, \ldots,0]} (z_a) &=&  \sum_{a=1}^{n} \left( z_a + \frac{1}{z_a} \right)~. \label{charfunds}
\eea
Using formula \eref{PEgen} and the expansion $- \log (1-x) = \sum_{k=1}^\infty x^k/k$, we can write the plethystic exponential as
\bea
\mathrm{PE}\: \left[ \chi^{B_{n}}_{[1,0, \ldots,0]}(z_l) \sum_{i=1}^{N_f} t_i  \right] &=& \frac{1}{ \prod_{i=1}^{N_f} \prod_{a=1}^{n} (1-t_i)(1-t_i z_a) \left(1-\frac{t_i}{ z_a} \right)}~, \nn \\
\mathrm{PE}\: \left[ \chi^{D_{n}}_{[1,0, \ldots,0]}(z_l) \sum_{i=1}^{N_f} t_i  \right] &=& \frac{1}{ \prod_{i=1}^{N_f} \prod_{a=1}^{n} (1-t_i z_a) \left(1-\frac{t_i}{ z_a} \right)}~.
\eea

\noindent The roots of the Lie algebras of $B_{n}$ and $D_{n}$ are respectively $\{ \pm L_a \pm L_b, \pm L_a\}$ (with $a \ne b$) and $\{ \pm L_a \pm L_b\}$ (with $a \ne b$).  Haar measures of special orthogonal groups can be written explicitly using Weyl's integration formula (see, {\em e.g.}, Section 26.2 of \cite{FH}):
\bea
\int_{B_{n}} \ud \mu_{B_{n}} &=& \frac{1}{(2 \pi i)^{n} n! 2^{n}} \oint_{|z_1|=1} \ldots \oint_{|z_{n}|=1}  \frac{ \ud z_1}{z_1} \ldots \frac{ \ud z_{n}}{z_{n}} \prod_{\alpha} \left(1- \prod_{l=1}^{n}z_l^{\alpha_l} \right), \nn \\ 
\int_{D_{n}} \ud \mu_{D_{n}} &=& \frac{1}{(2 \pi i)^{n} n! 2^{n-1}} \oint_{|z_1|=1} \ldots \oint_{|z_{n}|=1}  \frac{ \ud z_1}{z_1} \ldots \frac{ \ud z_{n}}{z_{n}} \prod_{\beta} \left(1- \prod_{l=1}^{n}z_l^{\beta_l} \right),~{}~{}~\label{haar}
\eea
where $\alpha$ and $\beta$ are respectively the roots of $B_{n}$ and $D_{n}$, and the notation $\alpha_l$ (resp. $\beta_l$) denotes the number in the $l$-th position of the root $\alpha$ (resp. $\beta$).
For reference, we shall give explicit examples for small values of $N_c$:
\bea 
\int_{SO(3)} \ud \mu_{SO(3)} &=& \frac{1}{2\times2 \pi i } \oint_{|z|=1}  \frac{ \ud z}{z} \left(1-\frac{1}{z}\right) \left(1-z\right)~, \nn \\
\int_{SO(4)} \ud \mu_{SO(4)} &=& \frac{1}{4\times(2 \pi i)^2 } \oint_{|z_1|=1} \frac{ \ud z_1}{z_1} \oint_{|z_2|=1} \frac{ \ud z_2}{z_2}  \left(1-\frac{z_1}{z_2}\right) \left(1-\frac{z_2}{z_1}\right) \left(1-\frac{1}{z_1 z_2}\right) (1-z_1 z_2)~, \nn \\
\int_{SO(5)} \ud \mu_{SO(5)} &=& \frac{1}{8\times(2 \pi i)^2 } \oint_{|z_1|=1} \frac{ \ud z_1}{z_1} \oint_{|z_2|=1} \frac{ \ud z_2}{z_2} \left(1-\frac{1}{z_1}\right) (1-z_1) \left(1-\frac{1}{z_2}\right) (1-z_2) \times \nn \\  &&  \left(1-\frac{z_1}{z_2}\right) \left(1-\frac{z_2}{z_1}\right) \left(1-\frac{1}{z_1 z_2}\right) (1-z_1 z_2)~. \label{exphaar}
\eea

As an example, we shall demonstrate how to calculate the generating function $g^{(3,SO(3))}$.  Putting the above together, we find that
\beq
g^{(3, SO(3))} (t_1, t_2, t_3) =  \frac{1}{2\times2 \pi i } \oint_{|z|=1}  \frac{ \ud z}{z} \frac{\left(1-\frac{1}{z}\right) \left(1-z\right)}{ \prod_{i=1}^3 \left[ (1-t_i)(1-t_i z) \left(1-\frac{t_i}{z} \right) \right]}
\eeq
Using the residue theorem with the poles $z = 0,\: t_1,\: t_2,\: t_3$ within the unit circle, we find that
\beq
g^{(3, SO(3))} (t_1, t_2, t_3) = \frac{1-t_1^2 \: t_2^2 \: t_3^2}{(1-t_1t_2t_3) \prod_{1\leq i \leq j \leq 3} (1 - t_i t_j)}~.
\eeq
Note that upon setting $t_1= t_2 = t_3 = t$, we recover formula \eref{HSnfnc} with $N_c=3$.  The latter is called the \emph{unrefined} generating function.  It suffices for our purposes to compute \emph{unrefined} generating functions (\emph{i.e.} setting all $t_i = t$).  Results are listed below:

\paragraph{The case of $N_c =3$.}  Let us compute unrefined generating functions for the $N_c =3$ theory.  We have
\beq
g^{(N_f, SO(3))} (t) =  \frac{1}{2\times2 \pi i } \oint_{|z|=1}  \frac{ \ud z}{z} \frac{\left(1-\frac{1}{z}\right) \left(1-z\right)}{\left[ (1-t)(1-t z) \left(1-\frac{t}{z} \right) \right]^{N_f}}.
\eeq
Using the residue theorem with the poles at $z=0$ and $z=t$, we find the following generating functions:
\bea 
g^{(1,SO(3))}(t) &=& \frac{1}{1-t^2} = 1+ t^2+ t^4+ t^6+ t^8+ t^{10}+\ldots~, \nn \\
g^{(2,SO(3))}(t) &=& \frac{1}{\left(1-t^2\right)^3} = 1+3 t^2+6 t^4+10 t^6+15 t^8+21 t^{10}+ \ldots~, \nn \\
g^{(3,SO(3))}(t) &=& \frac{1-t^6}{(1-t^2)^6(1-t^3)} = \frac{1+t^3}{(1-t^2)^6} \nn \\ 
&=& 1+6 t^2+t^3+21 t^4+6 t^5+56 t^6+21 t^7+126 t^8+56 t^9+252 t^{10}+\ldots~, \nn \\
g^{(4,SO(3))}(t) 
&=& \frac{1 + t^2 + 4 t^3 + t^4 + t^6}{(1-t^2)^9} \nn \\
&=& 1+10 t^2+4 t^3+55 t^4+36 t^5+220 t^6+180 t^7+714 t^8+660 t^9+1992 t^{10}+\ldots~, \nn \\
g^{(5,SO(3))}(t) 
&=& \frac{1 + 3 t^2 + 10 t^3 + 6 t^4 + 6 t^5 + 10 t^6 + 3 t^7 + t^9}{(1-t^2)^{12}} \nn \\  
&=& 1+15 t^2+10 t^3+120 t^4+126 t^5+680 t^6+855 t^7+3045 t^8+\ldots~,  \nn \\
g^{(6,SO(3))}(t) 
&=& \frac{1 + 6 t^2 + 20 t^3 + 21 t^4 + 36 t^5 + 56 t^6 + 36 t^7 + 21 t^8 + 
 20 t^9 + 6 t^{10} + t^{12}}{(1-t^2)^{15}} \nn \\
&=& 1+21 t^2+20 t^3+231 t^4+336 t^5+1771 t^6+2976 t^7+10521 t^8+\ldots~, \nn \\
g^{(7,SO(3))}(t) 
&=&  1+28 t^2+35 t^3+406 t^4+756 t^5+4060 t^6+8478 t^7+30975 t^8+\ldots~.
\eea 

\paragraph{The case of $N_c=4$.}  Let us compute generating functions for the $N_c =4$ theory:
\beq
g^{(N_f, SO(4))} (t) = \frac{1}{4\times(2 \pi i)^2 } \oint_{|z_1|=1} \frac{ \ud z_1}{z_1} \oint_{|z_2|=1} \frac{ \ud z_2}{z_2}  \frac{ \left(1-\frac{z_1}{z_2}\right) \left(1-\frac{z_2}{z_1}\right) \left(1-\frac{1}{z_1 z_2}\right) (1-z_1 z_2) }{\left[ (1-t z_1) \left(1-\frac{t}{z_1} \right) (1-t z_2) \left(1-\frac{t}{z_2} \right)  \right]^{N_f}}.
\eeq
Integrating along the contour $|z_2|=1$ enclosing the poles $z_2 =0,\: t$ and then along the contour $|z_1|=1$ enclosing the poles $z_1 =0,\: t$, we find that

\bea 
g^{(1,SO(4))}(t) &=& \frac{1}{1-t^2} = 1+ t^2+ t^4+ t^6+ t^8+ t^{10}+\ldots~, \nn \\
g^{(2,SO(4))}(t) &=& \frac{1}{\left(1-t^2\right)^3} = 1+3 t^2+6 t^4+10 t^6+15 t^8+21 t^{10}+\ldots~, \nn \\
g^{(3,SO(4))}(t) &=& \frac{1}{\left(1-t^2\right)^6} 
= 1+6 t^2+21 t^4+56 t^6+126 t^8+252 t^{10}+\ldots~, \nn \\
g^{(4,SO(4))}(t) &=& \frac{1-t^8}{(1-t^2)^{10} (1-t^4)} =  \frac{1+t^4}{\left(1-t^2\right)^{10}} \nn \\
&=& 1+10 t^2+56 t^4+230 t^6+770 t^8+2222 t^{10}+\ldots~, \nn \\
g^{(5,SO(4))}(t) &=& \frac{1+t^2+6 t^4+t^6+t^8}{\left(1-t^2\right)^{14}} \nn \\
&=& 1+15 t^2+125 t^4+750 t^6+3585 t^8+14427 t^{10}+\ldots~, \nn \\
g^{(6,SO(4))}(t) &=& \frac{1+3 t^2+21 t^4+20 t^6+21 t^8+3 t^{10}+t^{12}}{\left(1-t^2\right)^{18}} \nn \\
&=& 1+21 t^2+246 t^4+2051 t^6+13377 t^8+72030 t^{10}+\ldots~, \nn \\
g^{(7,SO(4))}(t) &=& \frac{1+6 t^2+56 t^4+126 t^6+210 t^8+126 t^{10}+56 t^{12}+6 t^{14}+t^{16}}{\left(1-t^2\right)^{22}} \nn \\
&=& 1+28 t^2+441 t^4+4900 t^6+41944 t^8+291648 t^{10}+\ldots~, \nn \\
g^{(8,SO(4))}(t) &=& \tiny{\frac{1+10 t^2+125 t^4+500 t^6+1310 t^8+1652 t^{10}+1310 t^{12}+500 t^{14}+125 t^{16}+10 t^{18}+t^{20}}{\left(1-t^2\right)^{26}} }\nn \\
&=& 1+36 t^2+736 t^4+10536 t^6+114696 t^8+1000728 t^{10}+\ldots~.
\eea 

\paragraph{The case of $N_c=5$.} Finally, let us compute generating functions for the $N_c =5$ theory:
\bea 
g^{(N_f, SO(5))} (t) &=& \frac{1}{8\times(2 \pi i)^2 } \oint_{|z_1|=1} \frac{ \ud z_1}{z_1} \oint_{|z_2|=1} \frac{ \ud z_2}{z_2} \times \nn \\
&& \frac{\left(1-\frac{1}{z_1}\right) (1-z_1) \left(1-\frac{1}{z_2}\right) (1-z_2) \left(1-\frac{z_1}{z_2}\right) \left(1-\frac{z_2}{z_1}\right) \left(1-\frac{1}{z_1 z_2}\right) (1-z_1 z_2)}{\left[ (1-t) \left(1-\frac{t}{z_1}\right) (1-t z_1) \left(1-\frac{t}{z_2}\right) (1-t z_2) \right]^{N_f}}~. \nn \\
\eea
As before, we obtain generating functions as follows:

\bea 
g^{(1,SO(5))}(t) &=& \frac{1}{1-t^2} = 1+ t^2+ t^4+ t^6+ t^8+ t^{10}+\ldots~, \nn \\
g^{(2,SO(5))}(t) &=& \frac{1}{\left(1-t^2 \right)^3} 
= 1+3 t^2+6 t^4+10 t^6+15 t^8+21 t^{10}+\ldots~, \nn \\
g^{(3,SO(5))}(t) &=& \frac{1}{\left(1-t^2 \right)^6} 
= 1+6 t^2+21 t^4+56 t^6+126 t^8+252 t^{10}+\ldots~, \nn \\
g^{(4,SO(5))}(t)&=& \frac{1}{\left(1-t^2 \right)^{10}}
= 1+10 t^2+55 t^4+220 t^6+715 t^8+2002 t^{10}+\ldots~, \nn \\
g^{(5,SO(5))}(t) 
&=& \frac{1-t^{10}}{(1-t^2)^{15}(1-t^5)} = \frac{1+t^5}{(1-t^2)^{15}} \nn \\
&=& 1+15 t^2+120 t^4+t^5+680 t^6+15 t^7+3060 t^8+120 t^9+11628 t^{10}+\ldots~, \nn \\
g^{(6,SO(5))}(t) &=& \frac{1+t^2+t^4+6 t^5+t^6+t^8+t^{10}}{(1-t^2)^{20}} \nn\\
&=& 1+21 t^2+231 t^4+6 t^5+1771 t^6+120 t^7+10626 t^8+1260 t^9+53130 t^{10}+\ldots~, \nn \\
g^{(7,SO(5))}(t) &=& \frac{1+3 t^2+6 t^4+21 t^5+10 t^6+15 t^7+15 t^8+10 t^9+21 t^{10}+6 t^{11}+3 t^{13}+t^{15}}{(1-t^2)^{25}} \nn \\
&=& 1+28 t^2+406 t^4+21 t^5+4060 t^6+540 t^7+31465 t^8+7210 t^9+201376 t^{10}+\ldots~, \nn \\
g^{(8,SO(5))}(t) 
&=& 1+36 t^2+666 t^4+56 t^5+8436 t^6+1800 t^7+82251 t^8+29800 t^9+658008 t^{10}+\ldots~, \nn \\
g^{(9,SO(5))}(t) &=& 1+45 t^2+1035 t^4+126 t^5+16215 t^6+4950 t^7+194580 t^8+99550 t^9+\ldots~.
\eea 

From the above examples, it is amusing to observe that
\begin{observation} \label{sodeg}
The generating functions for the $N_f \geq N_c$ theory can be written as
\beq
g^{N_f \geq N_c} (t) = \frac{P_k(t)}{(1-t^2)^{\dim \CM_{(N_f, SO(N_c))}}}~, \nn
\eeq
where $P_k(t)$ is a degree $k$ polynomial such that $P_k(1) \neq 0$ and 
$\dim \CM_{(N_f, SO(N_c))} -k $ is a constant for a given $N_c$.  This constant, which is $\frac{1}{2}N_c(N_c-1) = {N_c \choose 2}$, can be computed from the case of $N_f = N_c$, where the moduli space is a complete intersection, using \eref{HSnfnc}.
\end{observation}
\noindent As we have seen several examples in \cite{Gray}, this observation also applies for SQCD with $SU(2)$ gauge group.  
Later we shall establish a similar observation for SQCD with $Sp$ gauge group.  

\subsection{Character Expansions and Global Symmetries} \label{charexpso}
In the previous section, we have obtained the generating functions analytically for various $(N_f, SO(N_c))$ theories.  As we mentioned earlier, the coefficients of $t^k$ in $g^{(N_f, SO(N_c))}(t)$ is the number of independent GIOs at the $k$-th order of quarks.  We shall see in this section that this number is in fact {\em the dimension of some representation of the global symmetry} at that order.  Moreover, we shall see that the character expansion allows us to write down the generating function for {\em any} $(N_f, SO(N_c))$ theory in a very compact and enlightening way.

\paragraph {Terminology.} In order to avoid cluttered notation, henceforth we shall abuse terminology by referring to each character by its corresponding representation.

\paragraph{The $N_f< N_c$ theories.}  Let us take the simplest example: $N_f< N_c$. 
From \eref{gNf<NcSO}, we see that the character expansion for the case of $N_f<N_c$ is
\beq \label{sonf<nc}
g^{N_f < N_c} (t_1, \ldots t_{N_f}) = \PE \left[ t^2 [2,0, \ldots, 0] \right] = \sum_{k=0}^\infty \mathrm{Sym}^k [2,0, \ldots,0] t^{2k}~,
\eeq 
where the second equality follows from the basic property of the plethystic exponential which produces all possible symmetric products of the function on which it acts.  We emphasise that we use the fully refined generating function which is a function of $N_f$ variables, and so this expression depends on $N_f$ variables, not just one variable $t$.  We note the identity (c.f. (3.4) of \cite{Gray})
\bea
\mathrm{Sym}^k [2,0, \ldots,0] = \sum_{n_1, \ldots, n_{N_f} \geq 0} [2n_1, 2n_2, \ldots, 2n_{N_f-1}] ~ \delta \left(k - \sum_{j=1}^{N_f} jn_j \right)~.
\eea
Therefore, we have the character expansion
\beq \label{charexpNf<Nc}
g^{N_f < N_c}(t_1, \ldots t_{N_f}) =  \sum_{n_1, \ldots, n_{N_f} \geq 0} [2n_1, 2n_2, \ldots, 2n_{N_f-1}] ~t^{2a}~,
\eeq
where $a =  \sum_{j=1}^{N_f} jn_j $~.
\paragraph{Character expansion of an arbitrary $(N_f, SO(N_c))$ theory.} 
From \eref{sonf<nc}, we see that the basic building block of the GIOs in the $SO$ theory is the irreducible representation with 2 Young boxes which are symmetrised.  Any other irreducible representation is obtained by repeating this basic building block and, in fact, each of such an irreducible representation appears precisely once in the character expansion.  However, when baryons get involved in the theory, this observation is slightly modified.  We propose selection rules for the coefficients of the character expansion of $g^{(N_f , SO(N_c))}$ as follows:
\begin{itemize}
\item For $N_f > N_c$, the numbers located after the $N_c$-th from the left are zeros, since any product of $M$'s and $B$'s antisymmetrised on $N_c + 1$ (or more) upper or lower ßavour indices must vanish;
\item The numbers located in the $1^\text{st}, 2^\text{nd}, \ldots, (N_c - 1)$-th postitions from the left are even, whereas the number in the $N_c$-th position can be either even or odd.  The latter is due to the fact that the baryon transforms in the representation $[0,0,\ldots, 1_{N_c;L},0, \ldots, 0]$ of the $SU(N_f)$ global symmetry.
\end{itemize}
Hence, the character expansion of the generating function of any $(N_f, N_c)$ theory is
\beq \label{gencharexp}
g^{(N_f > N_c , SO(N_c))}(t_1, \ldots t_{N_f}) =  \sum_{n_1, \ldots, n_{N_c} \geq 0} [2n_1, 2n_2, \ldots, 2n_{N_c-1}, n_{N_c}, 0, \ldots,0] ~t^{b}~,
\eeq
where $b =  2\sum_{j=1}^{N_c-1} jn_j + n_{N_c} N_c $~.  For $N_f = N_c$, this formula goes through and has the form
\beq \label{charexpNf=Nc}
g^{N_f = N_c}(t_1, \ldots t_{N_f}) =  \sum_{n_1, \ldots, n_{N_c} \geq 0} [2n_1, 2n_2, \ldots, 2n_{N_c-1}] ~t^{b}~.
\eeq

\paragraph{A non-trivial check of the general character expansion \eref{gencharexp}.} We note that the dimension of the representation $[a_1,\ldots,a_{n-1}]$ of $SU(n)$ is given by the formula (see, \emph{e.g.}, (15.17) of \cite{FH}):
\beq
\dim~[a_1,\ldots,a_{n-1}] = \prod_{1 \leq i < j \leq n} \frac{(a_i + \ldots + a_{j-1})+j-i}{j-i}~. \label{dimformula}
\eeq
Applying this dimension formula to the representations in (\ref{gencharexp}) for various $(N_f, N_c)$ and summing the series into closed forms, we obtain the expressions which are in agreement of the earlier results.

As an example for the $B_{n}$ gauge group, let us consider $(N_f=5,N_c=3)$ theory.  Using formula (\ref{dimformula}), we find that
\begin{equation}
\ba{rcl}
\dim~[2n_1, 2n_2,  n_3, 0]
&=& \frac{1}{4!~3!~2!~1!}\times \\ \nn
&& (2n_1 + 1)(2n_1 + 2n_2 + 2)(2n_1 + 2n_2 + n_3 + 3)(2n_1 + 2n_2 + n_3 + 0 + 4)\times \\ \nn
&& (2n_2 + 1)(2n_2 + n_3 + 2)(2n_2 + n_3 + 0 + 3)\times \\ \nn
&& (n_3 + 1)(n_3+ 0 + 2)\times \\ \nn
&& (0 + 1) ~.  \label{dim53B}
\ea
\end{equation}
Replacing the representation in (\ref{gencharexp}) with this expression and summing over $n_1, n_2, n_3, n_4$, we recover the expression
\beq
g^{(5,SO(3))} = \frac{1-3 t+9 t^2-9 t^3+9 t^4-3 t^5+t^6}{(1-t)^{12} (1+t)^9}~. \nn
\eeq
As an example for the $D_{n}$ gauge group, let us consider $(N_f=5,N_c=4)$ theory.  We have the dimension formula
\begin{equation}
\ba{rcl}
\dim~[2n_1, 2n_2,  2n_3, n_4]
&=& \frac{1}{4!~3!~2!~1!}\times \\ \nn
&& (2n_1 + 1)(2n_1 + 2n_2 + 2)(2n_1 + 2n_2 + 2n_3 + 3)(2n_1 + 2n_2 + n_3 + n_4 + 4)\times \\ \nn
&& (2n_2 + 1)(2n_2 + 2n_3 + 2)(2n_2 + 2n_3 + n_4 + 3)\times \\ \nn
&& (2n_3 + 1)(2n_3+ n_4 + 2)\times \\ \nn
&& (n_4 + 1) ~.  \label{dim53D}
\ea
\end{equation}
Replacing the representation in (\ref{gencharexp}) with this expression and summing over $n_1, n_2, n_3, n_4$, we recover the expression
\beq
g^{(5,SO(4))} = \frac{1+t^2+6 t^4+t^6+t^8}{\left(1-t^2\right)^{14}}~. \nn
\eeq

\subsection{Counting Basic Generators of Gauge Invariants and Syzygies: the Plethystic Logarithm}
We will use the {\bf plethystic logarithm} to deduce the number of generators and constraints at
each order of quarks and antiquarks from the generating function
\cite{feng, forcella}. We recall the expression for the plethystic logarithm, $\PL$, the inverse function to
$\PE$, is
\begin{equation} \label{PLso}
\mathrm{PL}[g^{(N_f , SO(N_c))}(t_1,\ldots t_{N_f})] = \sum_{k=1}^{\infty} \frac{\mu(k)}{k} \log(g^{(N_f , SO(N_c))}(t_1^k, \ldots, t_{N_f}^k )) ~,
\end{equation}
where $\mu(k)$ is the M\"obius function.
The significance of the series expansion of the plethystic logarithm is stated in \cite{feng, forcella}: \emph{the first terms with plus sign give the basic generators while the first terms with the minus sign give the constraints between these basic generators.}  If the formula (\ref{PLso}) is an infinite series of terms with plus and minus signs, then the moduli space is not a complete intersection and the constraints in the chiral ring are not trivially generated by relations between the basic generators, but receive stepwise corrections at higher degree. These are the so-called {\bf higher syzygies}.  We shall demonstrate these facts below.

\paragraph{The case of $N_c =3$.} 
\beq \ba{rcl}
\PL[g^{(2,SO(3))}(t)] &=& 3t^2~, \nn \\
\PL[g^{(3,SO(3))}(t)] &=& 6 t^2+t^3-t^6~, \nn \\
\PL[g^{(4,SO(3))}(t)] &=& 10 t^2+4 t^3-4 t^5-10 t^6+15 t^8+20 t^9+ \ldots~, \nn \\
\PL[g^{(5,SO(3))}(t)] &=&15 t^2+10 t^3-24 t^5-55 t^6+15 t^7+225 t^8+330 t^9+\ldots~, \nn\\
\PL[g^{(6,SO(3))}(t)] &=&  21 t^2+20 t^3-84 t^5-210 t^6+120 t^7+1575 t^8+2604 t^9+\ldots~, \nn \\
\PL[g^{(7,SO(3))}(t)] &=& 28 t^2+35 t^3-224 t^5-630 t^6+540 t^7+7350 t^8+13720 t^9+\ldots~.
\ea \eeq

\paragraph{The case of $N_c =4$.} 
\beq \ba{rcl}
\PL[g^{(2,SO(4))}(t)] &=& 3t^2~, \nn \\
\PL[g^{(3,SO(4))}(t)] &=& 6t^2~, \nn \\
\PL[g^{(4,SO(4))}(t)] &=& 10 t^2 + t^4 - t^8~, \nn \\
\PL[g^{(5,SO(4))}(t)] &=& 15 t^2+5 t^4-5 t^6-15 t^8+24 t^{10}+30 t^{12}+\ldots~, \nn \\
\PL[g^{(6,SO(4))}(t)] &=&  21 t^2+15 t^4-35 t^6-99 t^8+504 t^{10}+245 t^{12}+\ldots~, \nn \\
\PL[g^{(7,SO(4))}(t)] &=& 28 t^2+35 t^4-140 t^6-441 t^8+4620 t^{10}-1330 t^{12}+\ldots~, \nn \\
\PL[g^{(8,SO(4))}(t)] &=& 36 t^2+70 t^4-420 t^6-1540 t^8+27300 t^{10}-32150 t^{12} + \ldots~, \nn \\
\PL[g^{(9,SO(4))}(t)] &=& 45 t^2+126 t^4-1050 t^6-4536 t^8+121464 t^{10}-267765 t^{12}+\ldots~. 
\ea \eeq

\paragraph{The case of $N_c =5$.} 
\beq \ba{rcl}
\PL[g^{(2,SO(5))}(t)] &=& 3 t^2~, \nn \\
\PL[g^{(3,SO(5))}(t)] &=& 6 t^2~, \nn \\
\PL[g^{(4,SO(5))}(t)] &=& 10 t^2~, \nn \\
\PL[g^{(5,SO(5))}(t)] &=& 15 t^2+t^5-t^{10}~, \nn \\
\PL[g^{(6,SO(5))}(t)] &=& 21 t^2+6 t^5-6 t^7-21 t^{10}+35 t^{12}-15 t^{14}+70 t^{15}-210 t^{17}+\ldots~, \nn \\
\PL[g^{(7,SO(5))}(t)] &=& 28 t^2+21 t^5-48 t^7+28 t^9-231 t^{10}+980 t^{12}-1668 t^{14}+3080 t^{15}+\ldots~, \nn \\
\PL[g^{(8,SO(5))}(t)] &=& 36 t^2+56 t^5-216 t^7+280 t^9-1596 t^{10}-120 t^{11}+11760 t^{12}-37620 t^{14}+\ldots~, \nn \\
\PL[g^{(9,SO(5))}(t)] &=& 45 t^2+126 t^5-720 t^7+1540 t^9-8001 t^{10}-1440 t^{11}+88200 t^{12}+495 t^{13}+\ldots~. 
\ea \eeq

\paragraph{Character expansion of the plethystic logarithm.} We emphasise that coefficients in plethystic logarithmic series are dimensions of representations of the $SU(N_f)$ global symmetry.  It is therefore possible to calculate character expansions of plethystic logarithms in a similar fashion to those of generating functions.  However, since we are interested in basic generators and basic constraints, only first few terms are significant for our purposes.  Consider an example of $\PL \left[ g^{(9, SO(4))} (t_i) \right]$.  The character expansion is 
\bea \label{PL94}
\PL \left[ g^{(9, SO(4))} (t_1,\ldots, t_9) \right] &=& [2,0,0,0,0,0,0,0] t^2 + [0,0,0,1,0,0,0,0] t^4 - [1,0,0,0,1,0,0,0] t^6 \nn \\
  &-& \left( \mathrm{Sym}^2 [0,0,0,1,0,0,0,0] - [2,0,0,0,0,1,0,0] \right) t^8 + \ldots~. \nn \\
\eea
For the basic generators of the GIOs, the coefficient of $t^2$ indicates that there are mesons at order 2 and the coefficient of $t^4$ indicates that there are baryons at order 4.  For the basic constraints, the coefficient of $t^6$ suggests that there is a relation between the basic generators at order 6 given by \eref{cons1} and the symmetric square in the coefficient of $t^8$ suggests that there is also a relation at order 8 given by \eref{cons2}.  However, we can see that this relation at order 8 receives a correction $-[2,0,0,0,0,1,0,0] $, which results from the product between the generator at order $t^2$ and the relation at order $t^6$.  This correction is the first in an infinite tower of relations that will not be dealt with here. Note that for the general $SO(N_c)$ theory, such a product is at order $N_c +4$, and so we see that such a correction at order 8 occurs \emph{only when} $N_c = 4$ but not for other values of $N_c$.  


\section{$Sp(N_c)$ SQCD with $N_f$ flavours}  \label{sp} \setall
Let us consider an $Sp(N_c)$ gauge theory\footnote{We shall use the notation where the rank of $Sp(n)$ is $n$ and $Sp(1)$ is isomorphic to $SU(2)$.  This is in agreement with the notation of \cite{Intriligator:1995ne}.} with $N_f$ flavours of matter in the fundamental $2N_c$ dimensional representation.  In the same way as before, we shall specify such a theory by $(N_f, Sp(N_c))$. Since the number of fundamentals must be even according to \cite{Witten:1982fp},  we take our matter content to be the quarks $Q^i_a$, with supermultiplet index $i = 1, \ldots, 2N_f$ and colour index $a = 1, \ldots, 2N_c$.  Thus, there is a total of $4N_cN_f$ chiral degrees of freedom from the quarks. Their quantum numbers are summarised in Table \ref{t:sp}.  The excellent reviews collecting this work are \cite{Terning:2003th, Terning:2006bq, Intriligator:1995ne}.  
\begin{table}[htdp]
\begin{center}
\begin{tabular}{|c||c|ccc|}
\hline
 &Gauge symmetry&{}&Global symmetry&  \\
& $Sp(N_c)$ & $SU(2N_f)$ &$U(1)_B$ &$U(1)_R$ \\
\hline \hline
$Q^i_a$ & $\tiny\yng(1)$ & $\tiny\yng(1)$ & 1& $\frac{N_f-1-N_c}{N_f}$ \\
\hline
\end{tabular}
\end{center}
\caption{{\sf The gauge and global symmetries of $Sp(N_c)$ SQCD with $N_f$ flavours and the quantum numbers of the chiral supermultiplets. }}
\label{t:sp}
\end{table}

\subsection{The $N_f \leq N_c$ Theories} 
At a generic point in the classical moduli space, the $Sp(N_c)$ gauge symmetry is broken to $Sp(N_c - N_f)$.  Since the dimension of $Sp(N)$ is $N (2N+ 1)$, there are
\beq
N_c (2N_c+1) - (N_c - N_f) (2N_c -2 N_f +1) =  4N_cN_f - 2N_f^2 +N_f\nn
\eeq 
broken generators.   Therefore, of the original $4N_cN_f$ chiral supermultiplets, only
\beq
4N_cN_f - \left[ 4N_cN_f - 2N_f^2 +N_f \right] = N_f(2N_f -1) \nn
\eeq
singlets are left massless.  Hence, the dimension of the moduli space of vacua is
\beq \label{dimnf<=nc}
\dim\left( \CM_{N_f \leq N_c} \right) =  N_f(2N_f -1) ~.
\eeq
We can describe these light degrees of freedom in a gauge invariant way by the mesons:
\beq \label{mesonSp}
\ba{ll}
M^{i j} = Q^i_a Q^j_b J^{ab} & \qquad \mbox{(meson)} ~,
\ea
\eeq
where the matrix $J = \mathbf{1}_{N_c} \otimes i \sigma_2$ is an invariant of $Sp(N_c)$.  We emphasise that the indices $i$ and $j$ are antisymmetric.   Therefore, the meson transforms in the global $SU(N_f)$ representation $\Lambda^2 ~[1,0,\ldots,0] =  [0, 1, \ldots, 0]$.  We note that for the $N_f \leq N_c$ theory, there are no relations (constraints) between mesons.  Phrasing this geometrically, and noting the dimension from \eref{dimnf<=nc}, we have that 
\begin{observation} \label{modSpnf<nc}
The classical moduli space  $\CM_{N_f \leq N_c}$ is freely generated: there are no relations among the generators.
The space $\mathcal{M}_{N_f \leq N_c}$ is, in fact, nothing but $\IC^{N_f(2N_f-1)}$.
\end{observation}
\noindent Using the plethystic programme, we can immediately write down the generating function of GIOs for $N_f \leq N_c$ as
\beq \label{gNf<NcSp}
g^{N_f \leq N_c} (t) = \PE \left[ t^2 \dim~[0, 1, \ldots, 0]  \right] = \PE \left[ N_f(2N_f-1) t^2 \right] = \frac{1}{(1-t^2)^{N_f(2N_f-1)}}~.
\eeq
We emphasise that this formula does not depend on the number of colours $N_c$. This expression is simply the Hilbert series for $\IC^{N_f(2N_f-1)}$, with weight 2 for each meson.

\subsection{The $N_f > N_c$ Theories} 
At a generic point in the moduli space,  the $Sp(N_c)$ gauge symmetry is broken completely and hence the number of remaining massless chiral supermultiplets ({\em i.e.}\ the dimension of the moduli space) is given by
\begin{equation} \label{vevSp}
\dim \left( \mathcal{M}_{N_f > N_c} \right) = 4N_f N_c - N_c (2N_c+1) ~.
\end{equation}
These light degrees of freedom can be parametrised by the mesons given by \eref{mesonSp}.  We refer to a discussion in \cite{Intriligator:1995ne} that there is no baryon, since the invariant tensor $\epsilon^{a_1 \ldots a_{2N_c}}$ decomposes into sums of products of the $J^{ab}$ and so baryons break up into mesons.  There is also a basic constraint between mesons due to the fact that any product of $M$Õs antisymmetrised on $2N_c + 1$ (or more) upper or lower flavour indices vanishes:
\beq \label{consSp}
\epsilon_{i_1 \ldots i_{2N_f}} M^{i_1 i_2} M^{i_3 i_4} \ldots M^{i_{2N_c +1}  i_{2N_c+2}} = 0~.
\eeq
The meson and the constraint \eref{consSp} transform respectively in the global $SU(2N_f)$ representations $[0, 1, \ldots, 0]$ and $[0, \ldots, 0, 1_{2N_c +2; L}, 0, \ldots, 0]$~.  They are respectively $2N_f \choose 2$ and $2N_f \choose 2N_c+2$ dimensional.

Although the moduli space $\CM_{N_f > N_c}$  is not freely generated, the special case $N_f = N_c + 1$ has a special property:
\begin{observation} \label{ciSp} 
The moduli space $\CM_{N_f = N_c+1}$ is a complete intersection and is, in fact, a single hypersurface in $\mathbb{C}^{(2N_c +1)(N_c+1)}$.
\end{observation} 
\noindent This is because the number of the basic generators (which is ${2N_c+2 \choose 2} = 2N_c^2+3N_c +1= (2N_c +1)(N_c+1)$) minus  the number of the basic constraints (which is 1) is equal to the dimension of the moduli space (which is $2N_c^2+3N_c$).  Observation \ref{ciSp} allows us to immediately write down the generating function for the $(N_c+1, Sp(N_c))$ theory by noting that there are $(2N_c +1)(N_c+1)$ mesonic generators of weight $t^2$, subject to a relation of weight $t^{2Nc+2}$ . 
\bea \label{nf=nc+1}
g^{(N_c+1, Sp(N_c))} &=& \left( 1-t^{2N_c+2} \right) \PE \left[ t^2~\dim [0,1,0, \ldots,0]  \right]  \nn\\
&=& \left( 1-t^{2N_c+2} \right) \PE \left[ (2N_c +1)(N_c+1) t^2  \right] \nn \\
&=& \frac{1-t^{2N_c+2}}{(1-t^2)^{(2N_c +1)(N_c+1)}} \nn \\
&=& \frac{1+t^2+t^4+\ldots+t^{2N_c}}{(1-t^2)^{2N_c^2+3N_c}}~.
\eea

\subsection{Character Expansions} 
Let us examine character expansions of generating functions of the $Sp$ theory.

\paragraph{The $N_f \leq N_c$ theories.}  Consider the simplest example: $N_f \leq N_c$.  From \eref{gNf<NcSp}, the character expansion is
\beq \label{spnf<nc}
g^{N_f\leq N_c} (t_1, \ldots, t_{2N_f}) = \PE \left[ t^2 [0, 1, \ldots, 0]  \right] = \sum_{k=0}^\infty \mathrm{Sym}^k [0,1,0,\ldots,0] t^{2k}~,
\eeq
where the second equality follows from the basic property of the plethystic exponential which produces all possible symmetric products of the function on which it acts.  We emphasise that we have used the fully refined generating function which is a function of $2N_f$ variables, and so this expression depends on $2N_f$ variables, not just one variable $t$.  
We note the identity (c.f. (3.4) of \cite{Gray})
\bea
\mathrm{Sym}^k [0,1, \ldots,0] = \sum_{n_1, \ldots, n_{2N_f} \geq 0} [0, n_2, 0, n_4, 0, \ldots, 0 , n_{N_f-2},0] ~ \delta \left(k - \sum_{j=1}^{N_f} 2 jn_{2j} \right)~.
\eea
Therefore, we have the character expansion
\beq \label{spcharexpNf<Nc}
g^{N_f \leq N_c}(t_1, \ldots t_{2N_f}) =  \sum_{n_1, \ldots, n_{2N_f} \geq 0}  [0, n_2, 0, n_4, 0, \ldots, 0 , n_{N_f-2},0] ~t^{\alpha}~,
\eeq
where $\alpha =  \sum_{j=1}^{N_f} 2jn_{2j}$~.

\paragraph{Character expansion of an arbitrary $(N_f, Sp(N_c))$ theory.} From \eref{spnf<nc}, we see that the basic building block of the GIOs in the $Sp$ theory is the irreducible representation with 2 Young boxes which are antisymmetrised.  Any other irreducible representation is built out of this basic building block and, in fact, each of such an irreducible representation appears precisely once in the character expansion.  We propose selection rules for the coefficients of the character expansion of $g^{(N_f , Sp(N_c))}$ as follows:
\begin{itemize}
\item Every number located in an odd position from the left is zero;
\item For $N_f > N_c+1$, the numbers located after the $2N_c$-th position from the left are zeros, since any antisymmetrisation on $2N_c + 1$ (or more) flavour indices yields a zero.
\end{itemize}
It follows that the character expansion for an arbitrary $(N_f, Sp(N_c))$ theory is 
\beq \label{genSp}
g^{(N_f, Sp(N_c))}(t_1, \ldots t_{2N_f}) = \sum_{n_2, n_4, \ldots, n_{2N_c} \geq 0}  [0, n_2, 0, n_4, 0, n_6, 0, \ldots, 0, n_{2N_c}, 0, ..., 0] ~t^\beta~,
\eeq
where $\beta= \sum_{j=1}^{N_c} 2j n_{2j}$.
We note that for $N_c =1$, formula \eref{genSp} becomes
\beq
g^{(N_f, Sp(1))}(t_1, \ldots t_{2N_f}) = \sum_{k = 0}^\infty [0, k , 0 \ldots, 0]~t^{2k}~.
\eeq
Note that this is also a character expansion of the $SU(2)$ SQCD with $N_f$ flavour (see formula (5.4) of \cite{Gray}).  Such an agreement is to be expected because of an isomorphism between $Sp(1)$ and $SU(2)$.

\subsection{Plethystic Exponentials and Molien--Weyl Formula}    Let us denote a basis for the dual space of the Cartan subalgebra by $\{ L_m \}_{m=1}^{N_c}$.  We choose $L_m = (0, \ldots,0, 1_{m;L},0, \ldots, 0)$, where the length of the tuple is $N_c$. The weights of the fundamental representation are $\{\pm L_m \}$.  With this choice of $L$'s, we find the character of the fundamental representation to be
\bea
\chi^{Sp(N_c)}_{[1,0, \ldots,0]}(z_l) = \sum_{m=1}^{N_c} \left( z_m + \frac{1}{z_m} \right)~.
\eea
The roots of the Lie algebra of $Sp(N_c)$ are $\omega =  \pm L_m \pm L_n$.  Therefore, the Haar measure is given by
\bea 
\int_{Sp(N_c)} \ud \mu_{Sp(N_c)} = \frac{1}{(2 \pi i)^{N_c} N_c! 2^{N_c}} \oint_{|z_1|=1} \ldots \oint_{|z_{N_c}|=1}  \frac{ \ud z_1}{z_1} \ldots \frac{ \ud z_{N_c}}{z_{N_c}} \prod_{\omega} \left(1- \prod_{l=1}^{N_c}z_l^{\omega_l} \right),~{}~{}~{}~{}\quad{}
\eea
where $\omega_l$ is the number in the $l$-th position of the root $\omega$. 
Similarly to the case of the $SO(N_c)$ gauge group, we have
\bea \label{genintSp}
g^{(N_f, Sp(N_c))} &=& \int_{Sp(N_c)} \ud \mu_{Sp(N_c)}\: \mathrm{PE}\: \left[ \chi^{Sp(N_c)}_{[1,0, \ldots,0]}(z_l) \sum_{i=1}^{2N_f} t_i  \right] ~. \nn \\
&=& \frac{1}{(2 \pi i)^{N_c} N_c! 2^{N_c}} \oint_{|z_1|=1} \ldots \oint_{|z_{N_c}|=1}  \frac{ \ud z_1}{z_1} \ldots \frac{ \ud z_{N_c}}{z_{N_c}} \frac{\prod_{\omega} \left(1- \prod_{l=1}^{N_c}z_l^{\omega_l} \right)}{ \prod_{i=1}^{2N_f} \prod_{m=1}^{N_c} (1-t_i z_m) \left(1-\frac{t_i}{ z_m} \right)}~.\nn \\
\eea
For reference, we shall list a few unrefined (i.e. $t_i = t$ for all $i=1,\ldots, 2N_f$) generating functions for the $Sp(2)$ SQCD.  Using the residue theorem twice with the poles at $0$ and $t$, we find that
\bea
g^{(1, Sp(2))} (t) &=& \frac{1}{1-t^2} \nn \\
&=& 1+t^2+t^4+t^6+t^8+t^{10}+\ldots~, \nn \\
g^{(2, Sp(2))}(t) &=& \frac{1}{\left(1-t^2\right)^6} \nn \\
&=& 1+6 t^2+21 t^4+56 t^6+126 t^8+252 t^{10}+\ldots~, \nn \\
g^{(3, Sp(2))}(t) &=& \frac{1+t^2+t^4}{\left(1-t^2\right)^{14}} \nn \\
&=& 1+15 t^2+120 t^4+679 t^6+3045 t^8+11508 t^{10}+\ldots~, \nn \\
g^{(4, Sp(2))}(t)  &=& \frac{1+6 t^2+21 t^4+28 t^6+21 t^8+6 t^{10}+t^{12}}{\left(1-t^2\right)^{22}} \nn \\
&=& 1+28 t^2+406 t^4+4032 t^6+30744 t^8+191736 t^{10}+\ldots~, \nn \\
g^{(5, Sp(2))}(t)  &=& \frac{1+15 t^2+120 t^4+470 t^6+1065 t^8+1377 t^{10}+1065 t^{12}+470 t^{14}+120 t^{16}+15 t^{18}+t^{20}}{\left(1-t^2\right)^{30}} \nn \\
&=& 1+45 t^2+1035 t^4+16005 t^6+186285 t^8+1739133 t^{10}+\ldots~. 
\eea
We note that these results are consistent with the character expansion \eref{genSp}.  As an example, let us consider the $(2, Sp(2))$ theory:
\bea
g^{(2, Sp(2))}(t) = \sum_{n_2, n_4 \geq 0} \dim~[0, n_2, 0] ~ t^{2n_2 + 4n_4} 
= \frac{{}_2F_1 (3, 4; 2; t^2)}{(1-t^4)} = \frac{(1 + t^2)}{(1 - t^2)^5 (1 - t^4)}
= \frac{1}{\left(1-t^2\right)^6}~, \nn \\
\eea
where the second equality follows from the dimension formula (3.18) of \cite{Gray}.  As before, we see that the method of summing dimensions of representations into a closed form provides a non-trivial check of formula \eref{genSp}. 

We can also make a similar proposition to Observation \ref{sodeg} that   
\begin{observation} \label{spdeg}
The generating functions for the $N_f \geq N_c + 1$ theory can be written as
\beq
g^{N_f \geq N_c+1} (t) = \frac{P_k(t)}{(1-t^2)^{\dim \CM_{(N_f, Sp(N_c))}}}~, \nn
\eeq
where $P_k(t)$ is a degree $k$ polynomial such that $P_k(1) \neq 0$ and
$\dim \CM_{(N_f, Sp(N_c))} -k $ is a constant for a given $N_c$.   This constant, which is $N_c(2N_c+1)$, can be computed from the case of $N_f = N_c+1$, where the moduli space is a complete intersection, using \eref{ciSp}.
\end{observation}



\subsection{Plethystic Logarithms}  Recall that the plethystic logarithm of the generating function $g^{(N_f, Sp(N_c))}$ is given by
\begin{equation} \label{PLsp}
\mathrm{PL}[g^{(N_f , Sp(N_c))}(t)] = \sum_{k=1}^{\infty} \frac{\mu(k)}{k} \log(g^{(N_f , Sp(N_c))}(t^k)) ~.
\end{equation}
We emphasise again that the first terms with plus sign give the basic generators, whereas the first terms with the minus sign give the constraints between these basic generators.  If the formula (\ref{PLsp}) is an infinite series of terms with plus and minus signs, then the moduli space is not a complete intersection.
We shall list a few results for the $Sp(2)$ SQCD:
\bea
\PL \left[ g^{(1, Sp(2))} (t) \right] &=&  t^2~, \nn \\
\PL \left[ g^{(2, Sp(2))} (t) \right] &=&  6 t^2~, \nn \\
\PL \left[ g^{(3, Sp(2))} (t) \right] &=& 15 t^2 - t^6~, \nn \\
\PL \left[ g^{(4, Sp(2))} (t) \right] &=& 28 t^2-28 t^6+63 t^8-36 t^{10}-378 t^{12}+1728 t^{14}+\ldots~, \nn \\
\PL \left[ g^{(5, Sp(2))} (t) \right] &=& 45 t^2-210 t^6+1155 t^8-2376 t^{10}-19800 t^{12}+\ldots~.
\eea
Take an example of $\PL \left[ g^{(4, Sp(2))} (t) \right]$.  We see that the term $28t^2$ indicates that there are 28 basic generators (mesons) at the order of 2 quarks, and the term $-28t^6$ indicates that there are 28 basic constraints (given by \eref{consSp}) at the order of 6 quarks. 

\paragraph{Character expansion of the plethystic logarithm.}  We can make a character expansion of the plethystic logarithm in a similar fashion as for the $SO$ theory.  Consider an example of $\PL \left[ g^{(4, Sp(2))}  \right]$.  The character expansion is 
\bea \label{PL42}
\PL \left[ g^{(4, Sp(2))} (t_1,\ldots, t_8) \right] = [0,2,0,0,0,0,0] t^2 - [0,0,0,0,0,1,0] t^6 + \ldots~.
\eea
The coefficient of $t^2$ indicates that there is one generator (meson) that transforms in the 28 dimensional  $[0,2,0,0,0,0,0]$ representation, and the coefficient of $t^6$ suggests that there is one relation between the mesons at order 6, given by \eref{consSp}, that transforms in the 28 dimensional  $[0,0,0,0,0,1,0]$ representation.

\section{An Orientifold Projection} \label{orientifold} \setall
Having the character expansion of the Hilbert Series for SQCD with all classical gauge groups, we can now turn to study relations between different theories. One natural relation arises from analogy to certain string theory backgrounds \cite{HW, Kutasov, BH} that include orientifolds \cite{HZ, HK}. In such backgrounds it is rather common that the gauge group reduces by a $\mathbb{Z}_2$ projection from a unitary gauge group to a symplectic or an orthogonal gauge group. We will now study the action of this $\mathbb{Z}_2$ on the generators and relations of the Hilbert Series. Any string theory background which embeds the $Sp$ and $SO$ gauge groups through an orientifold projection will have to act on irreducible representations of the global symmetry in the way specified in this section. We will henceforth refer to the $\mathbb{Z}_2$ action as an orientifold action without specifying the explicit brane or other construction.

We remind the reader that the quiver diagram of SQCD with the $SU$ gauge group \cite{Gray} can be drawn as in Figure \ref{fig:su}.  
\begin{figure}[htbp]
\begin{center}
\includegraphics[angle=0, scale=0.8]{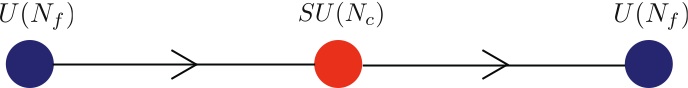}
\caption{{\sf The quiver diagram of $SU(N_c)$ SQCD with $N_f$ flavours.  The red node represents the $SU(N_c)$ gauge symmetry while the two blue nodes denote the global $U(N_f)$ symmetries. Each node gives rise to a $U(1)$ global symmetry, one of which is redundant.}} 
\label{fig:su}
\end{center}
\end{figure}
By introducing an orientifold, the $SU$ gauge group gets projected down to the $SO$ or $Sp$ gauge group, whereas the $U(N_f)_L\times U(N_f)_R$ flavor symmetry goes down to its diagonal $U(N_f)$ subgroup for the case of $SO$ gauge group and is enhanced to $U(2N_f)$ for the case of the $Sp$ gauge group. The orientifold action on the quiver diagram is to fold it along the red $SU(N_c)$ node, together with orientation reversal of the arrow in the quiver. As a result the red node becomes either $SO$ or $Sp$, depending on the projection, and the flavor symmetry either maps to itself for the case of $SO$ gauge group or is enhanced as stated above for the case of $Sp$ gauge group. The resulting quiver diagrams are shown in Figure \ref{fig:spso}.

\begin{figure}[htbp]
\begin{center}
\includegraphics[angle=0, scale=0.8]{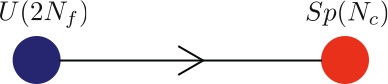} \hspace{2.5cm}
\includegraphics[angle=0, scale=0.8]{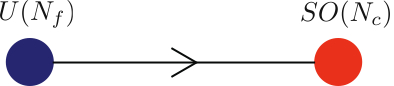}
\caption{{\sf {\bf Left diagram}: $Sp(N_c)$ SQCD with $N_f$ flavours as a quiver theory. The red node represents the $Sp(N_c)$ gauge symmetry, while the blue node denotes the global $U(2N_f)$ symmetries. {\bf Right diagram:}  $SO(N_c)$ SQCD with $N_f$ flavours as a quiver theory. The red node represents the $SO(N_c)$ gauge symmetry, while the blue node denotes the global $U(N_f)$ symmetries.  {\bf In each of these cases:} On the contrary to the $SU(N_c)$ SQCD, although the blue node give rise to a $U(1)$ factor,  the red node does not due to the orientifold projection.}}
\label{fig:spso}
\end{center}
\end{figure}

Having seen how an orientifold acts on the quiver diagram, one might ask what is the action of an orientifold on the global symmetries, basic generators of the GIOs, and basic constraints? 

\paragraph{An orientifold action on the global symmetry.}   The global $U(N_f) \times U(N_f)$ global symmetry of the $SU(N_c)$ theory gets projected down to its diagonal subgroup $U(N_f)$ for the $SO(N_c)$ gauge theory.  For the $Sp(N_c)$ gauge theory, as a result of the vanishing superpotential, the global symmetry further gets enhanced to $U(2N_f)$.  An antifundamental index in the $SU(N_c)$ theory becomes a fundamental index due to the orientation reversal in both $SO(N_c)$ and $Sp(N_c)$ theories. 

\paragraph{An orientifold action on the basic generators.}  A discussion on a similar problem is presented in \cite{Franco:2007ii}.  The mesons which transform in the bifundamental representation of $SU(N_f)_L \times SU(N_f)_R$ are projected down in two steps:  Firstly, the antifundamental index is turned into a fundamental index, and secondly, the resulting representation gets respectively symmetrised and antisymmetrised in the $SO(N_c)$ and $Sp(N_c)$ theories.  In the $SO(N_c)$ theory, the baryon and antibaryon get identifed by the orientifold projection and inherit the irreducible representation from the embedding of $U(N_f)$ inside $U(N_f) \times U(N_f)$.  On the other hand, in the $Sp(N_c)$ theory, a baryon breaks up into a product of mesons and stops being a generator of the chiral ring.  We summarise these results in Table \ref{basicgen}.

\begin{table}[htdp]
\begin{center}
\begin{tabular}{|c|c|c|c|}
\hline 
& $SU(N_c)$ SQCD & $Sp(N_c)$ SQCD & $SO(N_c)$ SQCD \\ \cline{2-4} 
Basic GIOs &  \multicolumn{3}{|c|}{Representation of the global symmetry} \\ \cline{2-4} 
& $SU(N_f)_L \times SU(N_f)_R$ & $SU(2N_f)$  & $SU(N_f)$ \\
\hline
 Meson & $[1,0, \ldots, 0; 0, \ldots, 0,1]$ & $[0, 1, 0, \ldots, 0]$ & $[2,0, \ldots, 0]$ \\
 Baryon &$[0,0,\ldots, 1_{N_c;L},0, \ldots, 0; 0,\ldots, 0]$ & * & $[0,0,\ldots, 1_{N_c;L},0, \ldots, 0]$ \\
 Antibaryon & $[0, \ldots,0 ; 0, \ldots, 1_{N_c;R},0 \ldots, 0]$ &* & ** \\
\hline
 \end{tabular}
\end{center}
\caption{{\sf The basic generators of GIOs for $SU$, $Sp$ and $SO$ SQCD and how they transform under the global symmetries. In the above, * indicates that in the $Sp$ theory baryons and antibaryons are simply products of mesons and stop being generators, and ** indicates that the antibaryon gets identified with the baryon such that we have one operator instead of two.}}
\label{basicgen}
\end{table}

\noindent Using these observations, we can immediately write down Hilbert Series for the freely generated moduli spaces in the $Sp$ and $SO$ theories starting from the one for $SU$ theory as follows:
\begin{displaymath}
\xymatrix{
\PE \left[ \mathrm{Sym}^2 [1, 0, \ldots, 0]_{SU(N_f)} t^2\right] = \PE \left[ [2, 0, \ldots, 0]_{SU(N_f)} t^2\right] \\
\PE \left[ [1,0, \ldots, 0; 0, \ldots, 0,1]_{SU(N_f)_L \times SU(N_f)_R}  t\: \tilde{t}\: \right] \ar[u]|{SO}  \ar[d]|{Sp} \\
\PE \left[ \Lambda^2 [1, 0, \ldots, 0]_{SU(2N_f)} t^2  \right]  = \PE \left[ [0, 1, 0, \ldots, 0]_{SU(2N_f)} t^2\right]  }
\end{displaymath}

\paragraph{An orientifold action on the basic constraints.}   As for the basic generators, the projection occurs in two steps:  The antifundamental index is first turned into a fundamental index, and the resulting symmetry then gets respectively symmetrised and antisymmetrised in the $SO(N_c)$ and $Sp(N_c)$ theories.  The results are summarised in Table \ref{basiccons}.

\begin{table}[htdp]
\begin{center}
\begin{tabular}{|c|c|c|c|}
\hline 
Type & $SU(N_c)$ SQCD & $Sp(N_c)$ SQCD & $SO(N_c)$ SQCD \\ \cline{2-4} 
of &  \multicolumn{3}{|c|}{Representation of the global symmetry} \\ \cline{2-4} 
relations& $SU(N_f)_L \times SU(N_f)_R$ & $SU(2N_f)$  & $SU(N_f)$ \\
\hline
$BB=$ & $[\quad 1_{N_c;L} \quad ; \quad 1_{N_c;R}\quad]  $ &  $\dagger$  & $\mathrm{Sym}^2 [\: 1_{N_c;L} \:]$ \\
 $M \ldots M$ & & & \\
 \hline
                     &$[\quad 1_{(N_c+1);L} \quad;\quad 1_{1;R}]$ & & \\ 
 $MB=MB$     &  &  $[\quad 1_{2N_c +2; L} \quad]$ & $[1_{1;L} \quad 1_{N_c+1;L} \quad]$ \\
                    & $[1_{1;L} \quad;\quad 1_{(N_c+1);R} \quad]$  & & \\
\hline
 \end{tabular}
\end{center}
\caption{{\sf The basic constraints between GIOs for $SU$, $Sp$ and $SO$ SQCD and how they transform under the global symmetries.  Only non-zero components of representations are presented. $\dagger$ indicates that in the $Sp$ theory, the relation $BB = M \ldots M$ provides us with no new information, as a baryon are simply a product of mesons. }}
\label{basiccons}
\end{table}

\section{A Geometric Aper\c{c}u} \label{apercu} \setall

In \cite{Gray} and the preceding sections, we have used the plethystic programme, the Molien--Weyl formula and the character expansion technique, to construct generating functions (Hilbert Series) which count GIOs in SQCD with \emph{any} classical gauge group.  In the following, we use Hilbert Series to extract a number of useful geometrical properties of moduli spaces.
We note, \emph{en passant}, that  there have been a number of studies of moduli spaces using techniques from computational algebraic geometry \cite{Gray, Gray:2005sr, Gray:2006jb, Gray:2008zs, fluxcomp}.

\subsection{Palindromic Numerator}  We have observed in many case studies before that the numerator of the generating function (Hilbert series) for SQCD is palindromic, {\em i.e.}\ it can be written in the form of a degree $k$ polynomial:
\begin{equation}
P_k(t) = \sum_{n=0} ^ k a_n t^n ~,
\end{equation}
with symmetric coefficients $a_{k-n} = a_n$.
A trivial modification of the rigorous proof given in Section 4.3 of \cite{Gray} for the $SU(N_c)$ SQCD tells us that this palindromic property holds in general for the $SO$ and $Sp$ SQCD:
\begin{theorem} \label{palin} Let $P_k(t)$ be a numerator of the generating function (Hilbert series) $g^{(N_f, SO(N_c))}(t)$ or $g^{(N_f, Sp(N_c))}(t)$ and suppose that $P_k(1) \ne 0$.
Then, $P_k(t)$ is palindromic.
\end{theorem}

\subsection{The SQCD vacuum Is Calabi-Yau} 
Similar situations were encountered in \cite{Gray, Forcella:2008bb}.  Due to a well-known theorem in commutative algebra called the Hochster--Roberts theorem\footnote{This theorem states that the invariant ring of a linearly reductive group acting on a regular ring is Cohen--Macaulay.}${}^,$\footnote{We are grateful to Richard Thomas for drawing our attention to this important theorem.}  \cite{hochster}, our coordinate rings of the moduli space are Cohen--Macaulay.
Therefore, as an immediate consequence of Theorem \ref{palin} and the Stanley theorem\footnote{This theorem states that the numerator to the Hilbert series of a graded Cohen--Macaulay domain $R$ is palindromic if and only if $R$ is Gorenstein.} \cite{stanley}, the chiral rings are also algebraically Gorenstein.   Since affine Gorenstein varieties means Calabi--Yau, we reach an important conclusion that $\CM_{(N_f ,N_c)}$ is, in fact, an affine Calabi--Yau cone over a weighted projective variety.  In brief,
\begin{observation}
The moduli spaces of the $SO(N_c)$ and $Sp(N_c)$ theories are Calabi-Yau.  
\end{observation}

\subsection{The SQCD Moduli Space Is Irreducible}
We start this subsection by noting that the irreducibility of moduli spaces is certainly not a feature of generic gauge theories; many reducible cases exist in the literature from very early studies of supersymmetric gauge theories.  Few recent ones are presented, for example, in \cite{Gray, Forcella:2008bb, Berenstein:2002ge}.  However, we shall see below that
\begin{observation}
The classical moduli spaces of SQCD with $SO$ and $Sp$ gauge groups are irreducible for all $N_f$ and $N_c$.
\end{observation}

As in \cite{Gray}, the moduli space (in the absense of a superpotential) of SQCD can be described by a symplectic quotient:
\beq 
\BC^n // G = \BC^n / G^c~,
\eeq
where $n = N_f N_c, \ 4N_fN_c$ for $G = SO(N_c),\ Sp(N_c)$ and $G^c$ denotes their complexifications, $G^c = SO(N_c, \BC),\ Sp(N_c, \BC)$.   Since $\BC^n$ is irreducible and $G^c$ is a continuous group, we expect the resulting quotient to be also irreducible\footnote{We are grateful to Alberto Zaffaroni for this point.}.

\section*{Acknowledgements}
We are indebted to Richard Thomas and Alberto Zaffaroni for instructive discussions.   
We also thank James Gray, Yang-Hui He and Vishnu Jejjala for a closely related collaboration.
N.M.~would like to express his gratitude to the following:
his family for the warm encouragement and support; Alexander Shannon,
William Rubens, and Fabian Spill for dicussions and useful comments;  
and, finally, the DPST Project and the Royal Thai Government for funding his research.

\end{document}